\pdfoutput=1
\documentclass[runningheads]{llncs}
\usepackage{graphicx}

\usepackage{hyperref}

\usepackage{cite}
\usepackage{amsmath}    
\usepackage{enumerate}  
\usepackage{rotating}
\usepackage{comment}
\usepackage{pifont}

\usepackage{color,soul}
\usepackage{float}
\usepackage[table,xcdraw]{xcolor}
\usepackage{multirow}
\usepackage{hhline}
\usepackage{arydshln}
\usepackage{booktabs}

\usepackage{xspace}
\newcommand{\cmark}{\ding{51}}%
\newcommand{\xmark}{\ding{55}}%
\newcommand*{\eg}{\emph{e.g.}\@\xspace}
\newcommand*{\ie}{\emph{i.e.}\@\xspace}
\newcommand*{\proposed}{\textbf{ImplicitVol}\@\xspace}
\newcommand*{\media}{PlaneInVol~\cite{yeung2021learning}\@\xspace}

\begin{document}
\title{\proposed: Sensorless 3D Ultrasound Reconstruction with Deep Implicit Representation}
%
\titlerunning{Sensorless 3D Ultrasound Reconstruction with Deep Implicit Representation}


\author{Pak-Hei Yeung\inst{1,2} \and
Linde Hesse\inst{1,2} \and
Moska Aliasi\inst{3} \and
Monique Haak\inst{4} \and
the INTERGROWTH-21$\textsuperscript{st}$ Consortium\inst{4} \and
Weidi Xie\inst{1,5}\thanks{These authors jointly supervised this work.} \and
Ana I.L. Namburete\inst{2}$^{\star}$}
%
\authorrunning{PH. Yeung et al.}
%
\institute{Department of Engineering Science, Institute of Biomedical Engineering, University of Oxford, Oxford, United Kingdom\\
\email{pak.yeung@pmb.ox.ac.uk}
\and
Oxford Machine Learning in NeuroImaging Lab, Department of Computer Science, University of Oxford, Oxford, United Kingdom
\and
Division of Fetal Medicine, Department of Obstetrics, Leiden University Medical Center, 2333 ZA Leiden, The Netherlands
\and
Nuffield Department of Women's and Reproductive Health, University of Oxford, Oxford, United Kingdom
\and
Visual Geometry Group, Department of Engineering Science, University of Oxford, Oxford, United Kingdom
}

\maketitle              

%

\begin{abstract}
The objective of this work is to achieve sensorless reconstruction of a 3D volume from a set of 2D freehand ultrasound images with deep \emph{implicit representation}.
In contrast to the conventional way 
that represents a 3D volume as a \emph{discrete} voxel grid,
we do so by parameterizing it as the zero level-set of a \emph{continuous} function,
{\em i.e.}~\emph{implicitly} representing the 3D volume as 
a mapping from the spatial coordinates to the corresponding intensity values.
Our proposed model, termed as \proposed, 
takes a set of 2D scans and their estimated locations in 3D as input,
jointly refining the estimated 3D locations and learning a full reconstruction of the 3D volume.
When testing on real 2D ultrasound images,
novel cross-sectional views that are sampled from \proposed show significantly better visual quality than those sampled from existing reconstruction approaches,
outperforming them by over 30\%~(NCC and SSIM), 
between the output and ground-truth on the 3D volume testing data.
The code will be made publicly available.



\end{abstract}

\begin{figure*}
\centering
\includegraphics[width=\textwidth]{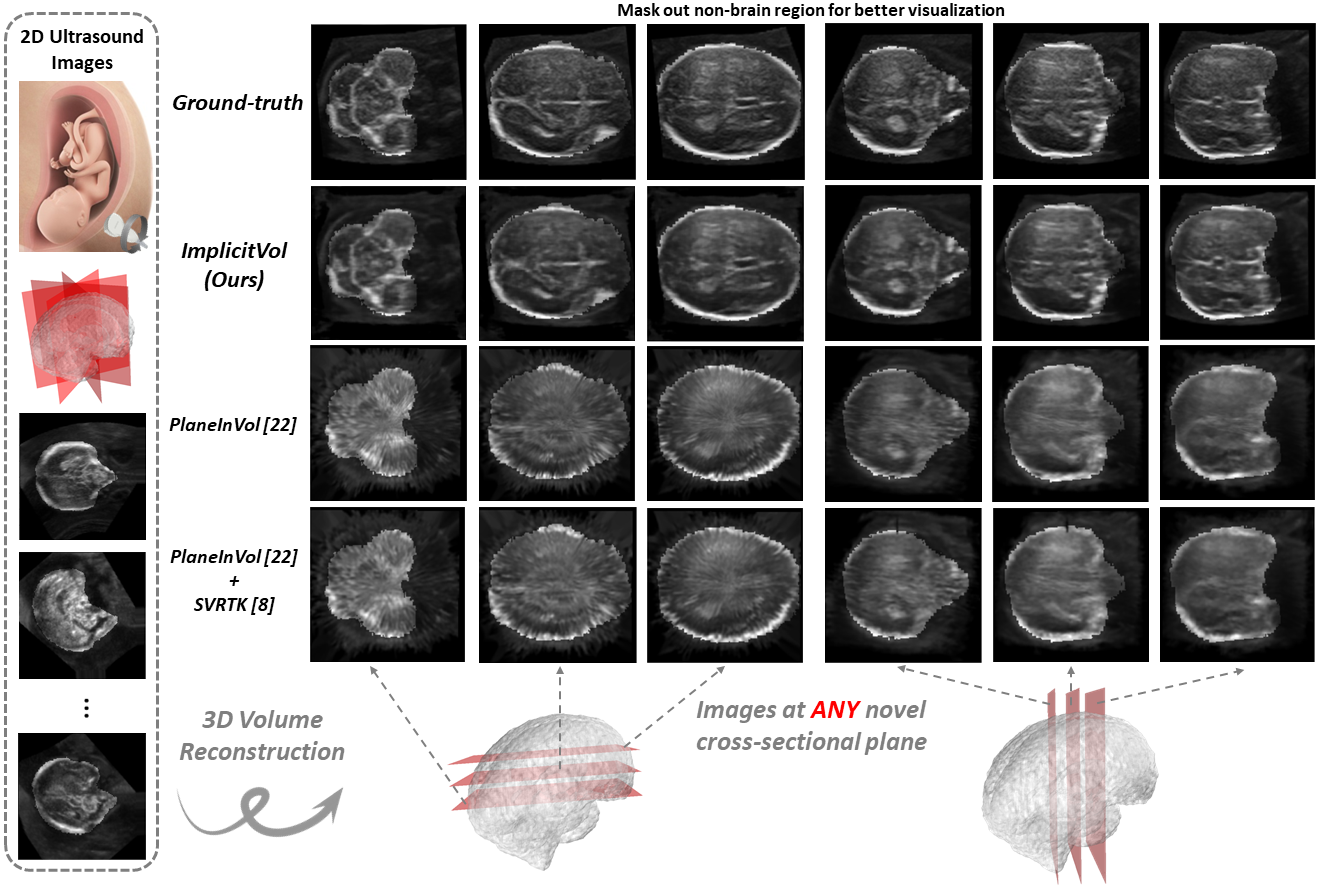}
\vspace{-20pt}
\caption{Visualization of 3D reconstruction from volume-sampled testing images by different approaches. 
Images sampled at different novel cross-sectional views,
masked by \cite{moser2019automated} for better visualization,
are presented and compared with the ground-truth. 
} 
\label{fig:synthetic}
\vspace*{-3pt}
\end{figure*}

%
%
%
\section{Introduction}
\label{sec:introduction}

Two-dimensional (2D) freehand ultrasonography
is a popular imaging tool 
for many clinical tests,
for example standard obstetric exams,
due to its cost-effectiveness, 
portability and real-time acquisition capabilities.
Despite its popularity,
since each 2D ultrasound scan only represents a cross-sectional view of the three-dimensional (3D) scanned structure~(Fig.~\ref{fig:synthetic}),
during scanning and visual analysis of the images, 
sonographers need to mentally reconstruct the 3D structures from the 2D scans.
This process requires extensive expertise and anatomical knowledge,
which may be limited in less developed regions, 
where medical professionals are insufficient~\cite{Benacerraf2002}.
In addition, 
since 2D freehand ultrasound scanning requires much manual operation, 
inter-operator variability may often be observed, 
and affect the accountability of the diagnosis.

Three-dimensional ultrasonography, on the other hand,
captures the whole structure as a 3D volumetric image during the scanning.
This leads to a range of advantages,
including shorter scanning and assessment times, 
more flexible offline and secondary examination, 
as well as providing richer diagnostic information
\cite{chen2001prenatal, pistorius2010grade, gonccalves2016three, duckelmann2010three}.
Nevertheless,
due to its more sophisticated hardware requirements and design,
a 3D ultrasound system is a lot bulkier and may cost ten times more than the 2D system,
limiting its use in practical scenario.
Here,
we capitalize on the advantages of 2D ultrasound,
by reconstructing 3D volumes from a set of freehand 2D ultrasound scans.

In the literature, 
as summarized in \cite{mohamed2019survey} and Section~\ref{sec:conventional},
early attempts tackled such a task by registering the 2D scans into the 3D volumes,
and explicitly performing interpolations in the resulting volumetric representation.
Despite the promising results achieved, 
multi-step reconstruction often suffers from different challenges,
for example, error accumulation due to the incorrect estimation from 2D scans to their corresponding 3D locations,
and limited resolution from the low tessellated grid.

In this paper,
we aim to tackle the aforementioned challenges
by parametrising the 3D volume as a deep neural network,
which jointly refines the 2D-to-3D registrations and 
learns full 3D reconstruction based on only a set of 2D scans. 
To the best of our knowledge,
our framework, \proposed,
is the first study to propose a \emph{genuine sensorless} 
(\ie in both training and inference)
3D reconstruction pipeline based on deep implicit representation.
We test \proposed on both volume-sampled and native freehand 2D ultrasound images.
Novel cross-sectional view images sampled from \proposed show significantly better visual quality and match (\ie more than 30\% improvement in structural similarity index) 
with the ground-truth
when compared to different baseline approaches
(Fig.~\ref{fig:synthetic}).
Although we only demonstrate the technique for 3D reconstruction of fetal brain ultrasound, 
the proposed approach is expected to be more general, 
and future work will aim to extend it to other modalities.

\section{Construction of 3D Representations}
\label{sec:preliminary}
In Section~\ref{sec:representation},
we briefly review the two ways of representing a 3D volume,
followed by the conventional 3D reconstruction approaches,
and our proposed approach with implicit representation in Section~\ref{sec:conventional}.

\subsection{Explicit and Implicit 3D Representations}
\label{sec:representation}
The very nature of a 3D volume is a one-to-one mapping from a set of 3D space positions (\ie 3D coordinates) to the corresponding intensity values in real world.
In general, there are two different ways for representing a 3D volume, 
either \emph{explicitly} or \emph{implicitly},
as described and compared in Table~\ref{table:preliminary} and the following sections:
\begin{table}[]
\centering
\setlength{\tabcolsep}{10pt}
\begin{tabular}{c|c|c}
    & \textbf{Explicit }      & \textbf{Implicit }      \\ \hline
\textit{continuity}       & discrete voxel grid         & continuous function     \\ \hline
\textit{memory-efficiency}       & lower       & higher    \\ \hline
\textit{resolution}     & defined by grid      & \begin{tabular}[c]{@{}l@{}} arbitrary \end{tabular} \\ \hline
\textit{\begin{tabular}[c]{@{}l@{}}gradient \& derivatives\end{tabular}} & \begin{tabular}[c]{@{}l@{}}limited by  discretization\end{tabular} & continuous \& well-defined           \\ \bottomrule     
\end{tabular}
\caption{Comparison between  the  explicit  and implicit representations for 3D volumes.}
\label{table:preliminary}
\vspace{-0.75cm}
\end{table}

\\[-8pt]
\par{\noindent \bf Explicit Representation.}
Conventionally, 
a 3D volume, $\mathbf{V} \in \mathcal{R}^{H \times W \times D \times C}$,
is represented \emph{discretely} and \emph{explicitly} as a tensor with height ($H$), width ($W$), depth ($D$), and intensity channels~($C$).
Most medical applications involving 3D volumes rely on using such representation. 
\\[-6pt]
    
\par{\noindent \bf Implicit Representation.}
As an alternative,
a 3D volume can also be represented as a zero level set of a {\em continuous} 
function parameterized by $\mathrm{\Theta}$.
Such {\em implicit} representation compresses the volumetric information
and encodes it as parameters of a model, 
for example a deep neural network,
that maps the 3D coordinates, $\mathbf{x} = (x, y, z)$, to intensities,
{\ie}~$F_\mathrm{\Theta} : \mathbf{x} \rightarrow c$.


\subsection{Conventional 3D Reconstruction Approaches}
\label{sec:conventional}

In the literature,
as summarized in \cite{mohamed2019survey},
early attempts on 2D-to-3D ultrasound reconstruction have been extensively built on the explicit representations, 
and can be summarised by the following steps: 

\begin{itemize}
\item[$\bullet$] First, 
the 3D location of each 2D ultrasound image is estimated,
where external sensor tracking is required at either the \emph{training}~\cite{prager2003sensorless, prevost2017deep} or \emph{inference}~\cite{daoud2015freehand, chung2017freehand} stage,
subject to errors caused by subjects' internal motion (\eg fetal movement).

\item[$\bullet$] Then,
a 3D volume, 
represented \emph{discretely} as a tensor of intensities,
is reconstructed by `registering' the localized 2D scans back to the 3D space,
with holes being interpolated.
However, such 2D-to-3D back-projections are often  
prone to errors, leading to artifacts,
thus requiring post-hoc corrections.


\item[$\bullet$] Finally,
approaches~\cite{chen2014reconstruction, moon20163d}
have been proposed to correct the aforementioned reconstruction artifacts,
based on kernel smoothing and denoising.
However the effect may be limited,
as the source of inaccuracy from the localization of the 2D images is unsolved,
which can be visualized in the last two rows of Fig.~\ref{fig:synthetic}.
\end{itemize}


\par
\begin{figure*}[t]
\centering
\includegraphics[width=\textwidth]{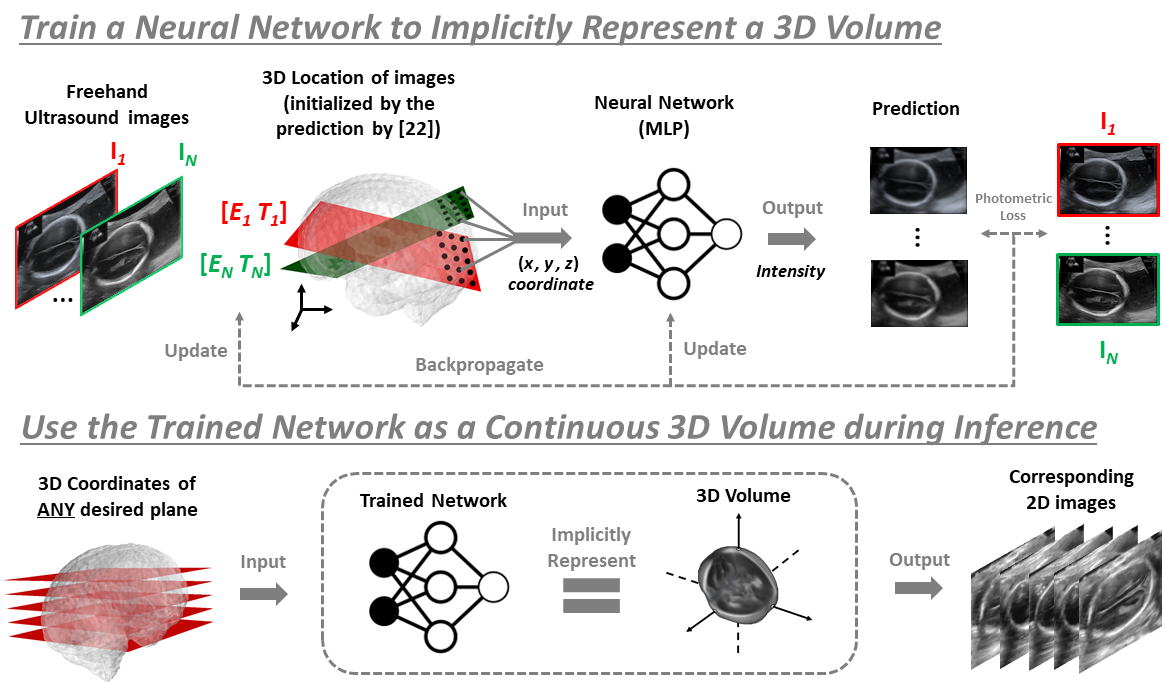}
\vspace{-20pt}
\caption{Pipeline of our proposed framework, \proposed. 
During training, 
a set of 2D freehand ultrasound images, $\{{\mathbf{I}_i}\}_{i=1}^N$,
and their estimated 3D location, $\{{\mathbf{E}_i}, {\mathbf{T}_i}\}_{i=1}^N$,
are used to train a deep neural network to implicitly represent the \emph{continuous} 3D volume from which 
$\{{\mathbf{I}_i}\}_{i=1}^N$ are acquired.
During inference, images at any planes can be obtained as output,
by feeding the corresponding grid coordinates to the network.
} 
\label{fig:pipeline}
\vspace{-10pt}
\end{figure*}

{\noindent \bf Contributions.}
In this paper, 
we parametrise a deep neural network for representing a 3D volume implicitly~(Fig.~\ref{fig:pipeline}).
Such a representation is continuous, 
and enables the querying of intensities at arbitrary spatial coordinates. 
With only a set of 2D scans available, 
it can produce jointly optimal 3D structures and 3D location estimations for these scans.

\section{Methods}
\label{sec:method}
In Section~\ref{sec:setup}, 
we first formulate the problem setting in this paper, 
namely, reconstructing a 3D volume from only a sparse set of 2D fetal brain scans with implicit representation.
Next, matching with the three conventional steps of 3D reconstruction summarized in Section~\ref{sec:conventional},
we introduce the corresponding components of \proposed,
namely sensorless 3D localization of 2D scans (Section~\ref{sec:localization}),
3D reconstruction with implicit representation (Section~\ref{sec:reconstruction})
and joint optimization (Section~\ref{sec:optimization}).
The whole pipeline is summarized in Fig.~\ref{fig:pipeline}.

\subsection{Problem Setup}
\label{sec:setup}
In general, 
we have a set of 2D ultrasound images, $\Pi = \{{\mathbf{I}_i}\}_{i=1}^N$, 
capturing $N$ different cross-sectional views of a fetal brain
at the corresponding 3D location,
parameterized by $\Lambda = \{{\mathbf{E}_i}, {\mathbf{T}_i}\}_{i=1}^N$,
with $\mathbf{E} = \{\theta_x, \theta_y, \theta_z\}$ being the 3D Euler angle and $\mathbf{T} = \{t_x, t_y, t_z\}$ denoting the translation. 
Our goal is to reconstruct the volume,
such that any 2D cross-sectional view of arbitrary resolution can be generated
by querying the corresponding 3D coordinates, $\mathbf{x} = (x, y, z)$.


Inspired by~\cite{sitzmann2019scene, mildenhall2020nerf},
we represent the volume as a continuous function,
parameterized by a multi-layer perceptron (MLP) representation network
$F_\mathrm{\Theta} : \mathbf{x} \rightarrow c$. 
The weights, $\mathrm{\Theta}$, are learned
by minimizing the the discrepancy between 
the actual and network-predicted intensities of $\{{\mathbf{I}_i}\}_{i=1}^N$,
when the 3D coordinates, 
$\mathbf{x}$, 
computed from the corresponding 
$\{{\mathbf{E}_i}, {\mathbf{T}_i}\}_{i=1}^N$,
are input to the network.




\subsection{Sensorless 3D Localization of 2D Scans}
\label{sec:localization}
We use \media for estimating the 3D locations,
$\Lambda = \{{\mathbf{E}_i}, {\mathbf{T}_i}\}_{i=1}^N$,
of the set of 2D ultrasound images, $\Pi = \{{\mathbf{I}_i}\}_{i=1}^N$.
Without using any external tracking,
\media is trained with a set of 2D slices,
sampled from the aligned 3D brain volumes,
and their locations in the 3D aligned space.
Despite being only trained with synthetic 2D images,
\media has demonstrated its generalizability when testing on real 2D freehand fetal brain images.
 
\subsection{3D Reconstruction with Implicit Representation}
\label{sec:reconstruction}
Conceptually, the idea is to \emph{store} the 3D volume in a MLP,
the weights of which are learned through a set of training data,
namely the 2D ultrasound images, $\Pi = \{{\mathbf{I}_i}\}_{i=1}^N$ 
and their pre-computed 3D locations, 
$\Lambda = \{{\mathbf{E}_i}, {\mathbf{T}_i}\}_{i=1}^N$,
detailed in Section~\ref{sec:localization}.

During training, 
we first derive the 3D coordinate, $\mathbf{x}_i^p \in \mathcal{R}^{3 \times 1}$, 
for pixel $p$ of the 2D ultrasound image, 
$\mathbf{I}_i \in \mathcal{R}^{H \times W}$,
from the estimated $\Lambda = \{{\mathbf{E}_i}, {\mathbf{T}_i}\}_{i=1}^N$:
\begin{equation}
\label{eq:coordinate_sum}
\mathbf{x}_i = \varepsilon (\mathbf{E}_i, \mathbf{T}_i)
\end{equation}
In practice,
this is achieved by first rotating 
the 3D coordinate of pixel $p$ of the reference $xy$ plane by $\mathbf{E}_i$,
and then translating it by ${\mathbf{T}_i}$.\\[-7pt]

\par{\noindent \bf Positional Encoding.}
Mapping each 3D coordinate, $\mathbf{x}$, 
to a higher dimensional space better represents the high frequency variation in the object's intensity and geometry
\cite{mildenhall2020nerf, rahaman2019spectral}.
Therefore, we encode $\mathbf{x}$ by the function $E : \mathcal{R} \rightarrow \mathcal{R}^{2L}$~\cite{mildenhall2020nerf}:
\begin{equation}
\label{eq:encoding}
E(n) = (\sin(2^0\pi n), \cos(2^0\pi n), ... , \sin(2^{L-1}\pi n), \cos(2^{L-1}\pi n))
\end{equation}
where $n$ are the normalized values (\ie from $-1$ to $1$) of each $x$, $y$ and $z$.
In the following sections,
``3D coordinate'' refers to the encoded coordinate,
$\mathbf{x} \in \mathcal{R}^{3 \times 2L}$. \\[-7pt]
%

\par{\noindent \bf Network Training.}
With the training set, $\{\mathbf{x}_i^p, \mathbf{I}_i^p\}\; \forall i, p$,
the weights, $\mathrm{\Theta}$, 
of the representation network, $F_\mathrm{\Theta}$, 
can be learned through normal back-propagation:
\begin{equation}
\label{eq:optimize}
\mathrm{\Theta}^{\ast }=\arg  \underset{\mathrm{\Theta} }{\min}\sum_{i,p}\mathcal{L}(F_\mathrm{\Theta}(\mathbf{x}_i^p), \mathbf{I}_i^p)
\end{equation}
where $\mathcal{L}$ is the photometric loss between observed and reconstructed 2D slices,
\ie structural similarity (SSIM) loss~\cite{wang2004image}.

\vspace{-3pt}
\subsection{Joint Optimization for Location Refinement}
\label{sec:optimization}
In practice,
the 3D locations, $\Lambda = \{{\mathbf{E}_i}, {\mathbf{T}_i}\}_{i=1}^N$, 
predicted by \media,
are imperfect due to prediction error.
Inspired by~\cite{wang2021nerfmm},
during training the network, $F_\mathrm{\Theta}$,
we update (\ie \emph{refine}) the pre-computed 3D locations, $\Lambda = \{{\mathbf{E}_i}, {\mathbf{T}_i}\}_{i=1}^N$,
simultaneously through joint optimization
which can be summarized as:
\begin{equation}
\label{equ:optimize_joint}
\mathrm{\Theta}^{\ast }, \mathrm{\Lambda}^{\ast } =
\arg  \underset{\mathrm{\Theta}, \mathrm{\Lambda} }{\min}
\; 
\mathcal{L}(F_\mathrm{\Theta}(E(\varepsilon(\Lambda))), \Pi)
\end{equation}
where $\Pi = \{{\mathbf{I}_i}\}_{i=1}^N$ and $\varepsilon(\cdot)$ and $E(\cdot)$ are from Eq.~\ref{eq:coordinate_sum} and \ref{eq:encoding}, respectively. \\[-6pt]

\par{\noindent \bf Inference.}
The trained representation network, $F_\mathrm{\Theta}$, 
represents a \emph{continuous} 3D fetal brain captured by the set of 2D images.
Any 2D cross-sectional view
at any resolution can be easily obtained as the output, 
by feeding the corresponding grid coordinates for the desired slice to the network,
as illustrated in the bottom half of Fig.~\ref{fig:pipeline}.

\section{Experimental Setup}
We test \proposed on both volume-sampled and native freehand 2D ultrasound fetal brain images,
and compare it with different baseline approaches. 
We use normalized cross-correlation (NCC)~\cite{yoo2009fast} and 
structural similarity index measure (SSIM)~\cite{wang2004image}, 
for appearance comparison,
as well as absolute difference between rotation angles and absolute distance between translations for location estimation comparison.
In Section~\ref{sec:data}, we introduce the datasets in this study,
followed by the experimental design in Section~\ref{sec:design} and
the implementation details in Section~\ref{sec:implement}.

\subsection{Dataset}
\label{sec:data}
Volume-sampled 2D images are generated by sampling planes from 
fifteen 3D ultrasound fetal brain volumes~($160^3$ voxels
),
collected at 20 weeks' gestational age.
The data were obtained as part of the INTERGROWTH-21st study~\cite{Papageorghiou2014}, 
which were collected using a Philips HD9 curvilinear probe at a 2–5 MHz wave frequency.

In addition, 
two videos of native freehand 2D brain scans with around 250 frames each,
collected at 20 weeks' gestational age at the Leiden University Medical Center using GE Voluson E10,
are used for qualitative analysis.

\subsection{Experimental Design}
\label{sec:design}
\par{\noindent \textbf{Volume-Sampled Images.}}
For each of the fifteen 3D volumes introduced in Section~\ref{sec:data},
($N \in \{128, 256\}$) 2D slices were sampled around the central axis of the brain non-uniformly, 
to simulate actual freehand acquisition by rotating the probe.
We conducted experiments on volume-sampled 2D images
because ideal ground-truth can be easily obtained
for quantitatively benchmarking different approaches. 
2D slices sampled at \textbf{new} cross-sectional views along the \emph{coronal}, \emph{sagittal} and \emph{axial} directions 
from both the original (\ie ground-truth) and reconstructed volumes by different approaches,
were analyzed.
The reconstructed volumes were rigidly aligned to the ground-truth volume for fair comparison
as rigid shift may be introduced to the volumes during the reconstruction.
The estimated 3D locations refined by \proposed were also compared to the ground-truth locations, and those predicted by other baseline approaches.\\[-6pt]


\par{\noindent \textbf{Real Images.}}
The two videos of real 2D freehand fetal brain ultrasound were acquired along the \emph{axial} direction.
Novel cross-sectional view images sampled from volumes 
reconstructed from different approaches were only {\em qualitatively} analyzed,
due to the lack of ground-truth 3D location information.

\subsection{Implementation Details}
\label{sec:implement}

\vspace{3pt}
\par{\noindent \textbf{\proposed.}}
Our representation network, $F_\mathrm{\Theta}$, is a 5-layer MLP,
with the hidden layer dimension of 128
and SIREN~\cite{sitzmann2020implicit} as the activation function.
$L$, from Eq.~\ref{eq:encoding}, was set to 10 and 
we initialized the set of 3D locations, $\Lambda = \{{\mathbf{E}_i}, {\mathbf{T}_i}\}_{i=1}^N$,
by the estimated locations predicted by \media.
The learning and decay rates followed those adopted in~\cite{wang2021nerfmm}. 
A representation network, $F_\mathrm{\Theta}$, 
was trained for one set of images
for 10000 epochs to represent one 3D volume. \\[-6pt]

\par{\noindent \textbf{\media.}}
With the set of 2D ultrasound images, $\Pi = \{{\mathbf{I}_i}\}_{i=1}^N$, 
and the corresponding 3D locations,
$\Lambda = \{{\mathbf{E}_i}, {\mathbf{T}_i}\}_{i=1}^N$,
predicted by \media,
$\mathbf{V} \in \mathcal{R}^{H \times W \times D}$
was explicitly reconstructed
by interpolating the intensity at each voxel 
by inverse distance weighted average 
from the 20 nearest pixel of $\Pi = \{{\mathbf{I}_i}\}_{i=1}^N$. \\[-6pt]

\par{\noindent \textbf{\media + SVRTK~\cite{kuklisova2012reconstruction}.}}
Slice to volume registration is well studied for super resolution reconstruction of motion-corrupted MRI.
We implemented SVRTK~\cite{kuklisova2012reconstruction},
designed for fetal brain MRI motion correction,
to the \textbf{\media interpolated} volume
to verify if using technique developed for a similar task in a different modality (\ie MRI) may help in our problem setting.

\label{sec:exp}


\section{Results and Discussion}

\label{sec:result}
\begin{figure*}[t]
\centering
\includegraphics[width=\textwidth]{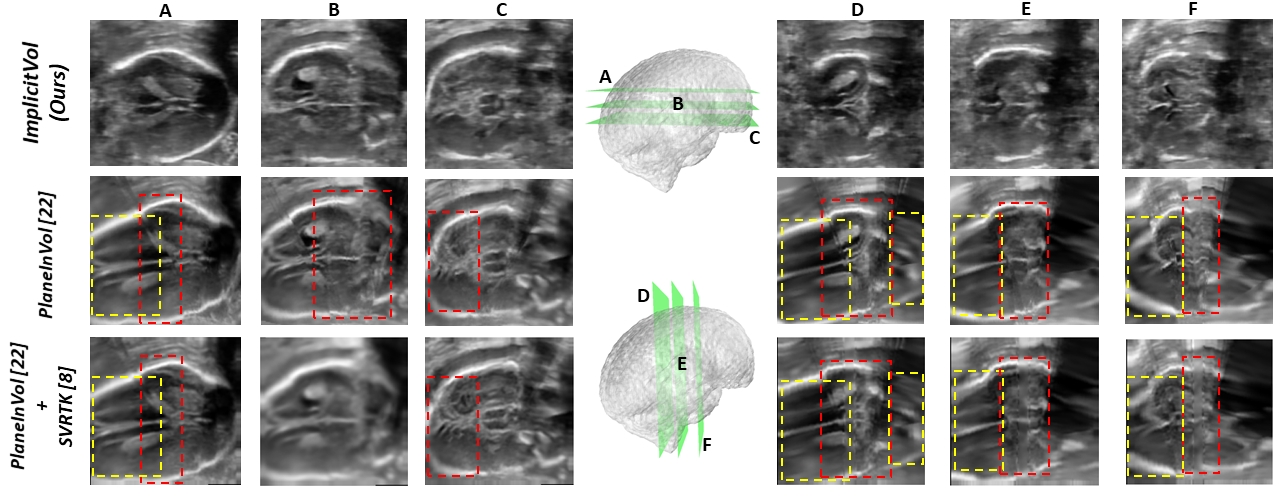}
\caption{Results of 3D reconstruction from native freehand 2D ultrasound. 
Novel view images sampled from different planes from volumes reconstructed by different approaches are presented. 
\proposed, shows better visual quality in under-sampled region (\textcolor{yellow}{yellow} boxes) and is more robust against inaccurate position estimation (\textcolor{red}{red} boxes).
} 
\label{fig:video}
\vspace*{-2.5mm}
\end{figure*}

The results of all the experiments are presented in Table~\ref{table:result}, 
with qualitative examples shown in Fig.~\ref{fig:video}.
In Section~\ref{sec:result_synthetic} and \ref{sec:result_real}, 
we analyze the results of volume-sampled and native freehand 2D ultrasound images, respectively.
\vspace*{-5mm}
\begin{table}[t]
\centering
\fontsize{6}{9}\selectfont
\begin{tabular}{|c|c|c|c|c|c|c|c|c|c|c|}
\hline
                                                                            &                                                                                     &                                                                                       & \multicolumn{2}{c|}{\textbf{Axial}}                                                                                                     & \multicolumn{2}{c|}{\textbf{Coronal}}                                                                                                     & \multicolumn{2}{c|}{\textbf{Sagittal}}                                                                                                     & \multicolumn{2}{c|}{\textbf{Location}}                                                                                                                     \\ \cline{4-11} 
\multirow{-2}{*}{\textbf{Approach}}                                                                                          & \multirow{-2}{*}{\textbf{\begin{tabular}[c]{@{}c@{}}No. of \\ Images\end{tabular}}} & \multirow{-2}{*}{\textbf{\begin{tabular}[c]{@{}c@{}} Jointly\\ Optimize\end{tabular}}} & \textbf{\begin{tabular}[c]{@{}c@{}}NCC\\ $\uparrow$\end{tabular}}     & \textbf{\begin{tabular}[c]{@{}c@{}}SSIM\\ $\uparrow$\end{tabular}}    & \textbf{\begin{tabular}[c]{@{}c@{}}NCC\\ $\uparrow$\end{tabular}}     & \textbf{\begin{tabular}[c]{@{}c@{}}SSIM\\ $\uparrow$\end{tabular}}    & \textbf{\begin{tabular}[c]{@{}c@{}}NCC\\ $\uparrow$\end{tabular}}     & \textbf{\begin{tabular}[c]{@{}c@{}}SSIM\\ $\uparrow$\end{tabular}}    & \textbf{\begin{tabular}[c]{@{}c@{}}Angle\\ (rad)$\downarrow$\end{tabular}} & \textbf{\begin{tabular}[c]{@{}c@{}}Distance\\ (pixel)$\downarrow$\end{tabular}} \\ \hline
(a) PlaneInVol\cite{yeung2021learning}                                                                                       & 128                                                                                 & -                                                                                     & \begin{tabular}[c]{@{}c@{}}0.569\\ $\pm$0.131\end{tabular}          & \begin{tabular}[c]{@{}c@{}}0.374\\ $\pm$0.074\end{tabular}          & \begin{tabular}[c]{@{}c@{}}0.563\\ $\pm$0.163\end{tabular}          & \begin{tabular}[c]{@{}c@{}}0.384\\ $\pm$0.071\end{tabular}          & \begin{tabular}[c]{@{}c@{}}0.485\\ $\pm$0.159\end{tabular}          & \begin{tabular}[c]{@{}c@{}}0.421\\ $\pm$0.098\end{tabular}          & \begin{tabular}[c]{@{}c@{}}0.237\\ $\pm$0.148\end{tabular}                & \begin{tabular}[c]{@{}c@{}}8.34\\ $\pm$5.32\end{tabular}                       \\ \hline
(b) PlaneInVol\cite{yeung2021learning}                                                                                       & 256                                                                                 & -                                                                                     & \begin{tabular}[c]{@{}c@{}}0.584\\ $\pm$0.129\end{tabular}          & \begin{tabular}[c]{@{}c@{}}0.387\\ $\pm$0.077\end{tabular}          & \begin{tabular}[c]{@{}c@{}}0.579\\ $\pm$0.162\end{tabular}          & \begin{tabular}[c]{@{}c@{}}0.392\\ $\pm$0.071\end{tabular}          & \begin{tabular}[c]{@{}c@{}}0.501\\ $\pm$0.158\end{tabular}          & \begin{tabular}[c]{@{}c@{}}0.432\\ $\pm$0.100\end{tabular}          & \begin{tabular}[c]{@{}c@{}}0.235\\ $\pm$0.149\end{tabular}                & \begin{tabular}[c]{@{}c@{}}8.49\\ $\pm$5.50\end{tabular}                       \\ \hline
\begin{tabular}[c]{@{}c@{}}(c) PlaneInVol\cite{yeung2021learning}\\ $\;\;\;\;$+ SVRTK\cite{kuklisova2012reconstruction}\end{tabular} & 128                                                                                 & -                                                                                     & \begin{tabular}[c]{@{}c@{}}0.580\\ $\pm$0.129\end{tabular}          & \begin{tabular}[c]{@{}c@{}}0.379\\ $\pm$0.071\end{tabular}          & \begin{tabular}[c]{@{}c@{}}0.570\\ $\pm$0.168\end{tabular}          & \begin{tabular}[c]{@{}c@{}}0.384\\ $\pm$0.072\end{tabular}          & \begin{tabular}[c]{@{}c@{}}0.495\\ $\pm$0.157\end{tabular}          & \begin{tabular}[c]{@{}c@{}}0.423\\ $\pm$0.093\end{tabular}          & \begin{tabular}[c]{@{}c@{}}0.237\\ $\pm$0.148\end{tabular}                & \begin{tabular}[c]{@{}c@{}}8.34\\ $\pm$5.32\end{tabular}                       \\ \hline
\begin{tabular}[c]{@{}c@{}}(d) PlaneInVol\cite{yeung2021learning}\\ $\;\;\;\;$+ SVRTK\cite{kuklisova2012reconstruction}\end{tabular} & 256                                                                                 & -                                                                                     & \begin{tabular}[c]{@{}c@{}}0.590\\ $\pm$0.127\end{tabular}          & \begin{tabular}[c]{@{}c@{}}0.389\\ $\pm$0.074\end{tabular}          & \begin{tabular}[c]{@{}c@{}}0.581\\ $\pm$0.171\end{tabular}          & \begin{tabular}[c]{@{}c@{}}0.392\\ $\pm$0.073\end{tabular}          & \begin{tabular}[c]{@{}c@{}}0.506\\ $\pm$0.155\end{tabular}          & \begin{tabular}[c]{@{}c@{}}0.434\\ $\pm$0.094\end{tabular}          & \begin{tabular}[c]{@{}c@{}}0.235\\ $\pm$0.149\end{tabular}                & \begin{tabular}[c]{@{}c@{}}8.49\\ $\pm$5.50\end{tabular}                       \\ \hline
\\ \hline
\multicolumn{1}{|l|}{(e) ImplicitVol}                                                                                        & 128                                                                                 & \xmark                                                                                & \begin{tabular}[c]{@{}c@{}}0.591\\ $\pm$0.137\end{tabular} &           \begin{tabular}[c]{@{}c@{}}0.428\\ $\pm$0.084\end{tabular} & \begin{tabular}[c]{@{}c@{}}0.584\\ $\pm$0.168\end{tabular}      & \begin{tabular}[c]{@{}c@{}}0.437\\ $\pm$0.067\end{tabular}  & \begin{tabular}[c]{@{}c@{}}0.511\\ $\pm$0.159\end{tabular}  & \begin{tabular}[c]{@{}c@{}}0.479\\ $\pm$0.092\end{tabular}  & \begin{tabular}[c]{@{}c@{}}0.237\\ $\pm$0.148\end{tabular}                & \begin{tabular}[c]{@{}c@{}}8.34\\ $\pm$5.32\end{tabular}                       \\ \hline
\multicolumn{1}{|l|}{(f) ImplicitVol}                                                                                        & 128                                                                                 & \cmark                                                                                & \begin{tabular}[c]{@{}c@{}}0.719\\ $\pm$0.096\end{tabular}          & \begin{tabular}[c]{@{}c@{}}0.535\\ $\pm$0.084\end{tabular}          & \begin{tabular}[c]{@{}c@{}}0.719\\ $\pm$0.116\end{tabular}          & \begin{tabular}[c]{@{}c@{}}0.534\\ $\pm$0.081\end{tabular}          & \begin{tabular}[c]{@{}c@{}}0.673\\ $\pm$0.117\end{tabular}          & \begin{tabular}[c]{@{}c@{}}0.581\\ $\pm$0.069\end{tabular}          & \begin{tabular}[c]{@{}c@{}}0.149\\ $\pm$0.090\end{tabular}                & \begin{tabular}[c]{@{}c@{}}6.73\\ $\pm$4.82\end{tabular}                       \\ \hline
\multicolumn{1}{|l|}{(g) ImplicitVol}                                                                                          & 256                                                                                 & \cmark                                                                                & \textbf{\begin{tabular}[c]{@{}c@{}}0.747\\ $\pm$0.096\end{tabular}} & \textbf{\begin{tabular}[c]{@{}c@{}}0.569\\ $\pm$0.085\end{tabular}} & \textbf{\begin{tabular}[c]{@{}c@{}}0.751\\ $\pm$0.098\end{tabular}} & \textbf{\begin{tabular}[c]{@{}c@{}}0.570\\ $\pm$0.080\end{tabular}} & \textbf{\begin{tabular}[c]{@{}c@{}}0.701\\ $\pm$0.116\end{tabular}} & \textbf{\begin{tabular}[c]{@{}c@{}}0.606\\ $\pm$0.072\end{tabular}} & \textbf{\begin{tabular}[c]{@{}c@{}}0.137\\ $\pm$0.097\end{tabular}}       & \textbf{\begin{tabular}[c]{@{}c@{}}6.32\\ $\pm$4.61\end{tabular}}              \\ \hline
\end{tabular}
\vspace*{1mm}
\caption{Evaluation results (mean $\pm$ standard deviation) of different approaches on volume-sampled 2D images. 
$\uparrow$ indicates higher values being more accurate, vice versa.
}
\label{table:result}
\end{table}


\vspace{8pt}
\subsection{Volume-Sampled Images}
\label{sec:result_synthetic}
A few conclusions can be drawn from the results presented in Table~\ref{table:result} and Fig.~\ref{fig:synthetic}.

\emph{Firstly},
The novel view images sampled from our proposed approach, \proposed (\textbf{rows e-g}),
showed a better match with the corresponding ground-truth as suggested by the higher NCC and SSIM values,
compared to the conventional approaches~(\textbf{rows a-d}).
Performance was improved by more than 30\%,
for all \emph{coronal}, \emph{sagittal} and \emph{axial} directions,
which can be further verified by the qualitative examples shown in Fig.~\ref{fig:synthetic}


\emph{Secondly},
while comparing \textbf{row e} to \textbf{row f}, 
updating the estimated 3D locations through the joint optimization in \proposed  
led to a significant boost of performance on visual quality
as well as more accurate estimations for localising the 2D images in the 3D space.
Note that, 
such refinement requires no extra supervision cost,
which manifests \proposed 's additional potential in slice-to-volume registration of ultrasound

\emph{Thirdly},
a larger training set (\ie 128 to 256) led to better performance (\textbf{row f} to \textbf{row g}).
Thanks to the real-time acquisition capability of 2D ultrasound,
acquiring hundreds of images in one scan is easily achievable in practice.

\subsection{Native Freehand Images}
\label{sec:result_real}
As shown by the native freehand 2D ultrasound results presented in Fig.~\ref{fig:video},
images sampled from \proposed
showed better visual quality at motion-corrupted regions (\textcolor{red}{red} boxes),
thanks to the localization refinement achieved by the joint optimization.
\proposed also performed better at under-sampled regions (\textcolor{yellow}{yellow} boxes),
where the baseline approaches reconstructed misleading results
due to the inaccuracy caused by extrapolation from the 
spatially distant neighbours.




\section{Conclusion}
In summary, we investigate on
sensor-free 3D ultrasound reconstruction from a sparse set of 2D images with deep implicit representation.
Our proposed framwork, \proposed, 
demonstrates superior performance,
in terms of 
the quality of the sampled images
as well as 
the refinement of the 3D localization,
when compared to other baseline approaches.
\proposed may facilitate more standardized analysis for 2D ultrasound sequence,
which may lead to better and more efficient diagnosis and assessment,
particularly in settings where only 2D probes are available but 3D assessment would be beneficial.




\bibliographystyle{splncs04}
\bibliography{ref}

\begin{thebibliography}{10}
\providecommand{\url}[1]{\texttt{#1}}
\providecommand{\urlprefix}{URL }
\providecommand{\doi}[1]{https://doi.org/#1}

\bibitem{Benacerraf2002}
Benacerraf, B.R.: Three-dimensional fetal sonography. Journal of Ultrasound in
  Medicine  \textbf{21}(10),  1063--1067 (2002).
  \doi{https://doi.org/10.7863/jum.2002.21.10.1063},
  \url{https://onlinelibrary.wiley.com/doi/abs/10.7863/jum.2002.21.10.1063}

\bibitem{chen2001prenatal}
Chen, M.L., Chang, C.H., Yu, C.H., Cheng, Y.C., Chang, F.M.: Prenatal diagnosis
  of cleft palate by three-dimensional ultrasound. Ultrasound in medicine \&
  biology  \textbf{27}(8),  1017--1023 (2001)

\bibitem{chen2014reconstruction}
Chen, X., Wen, T., Li, X., Qin, W., Lan, D., Pan, W., Gu, J.: Reconstruction of
  freehand 3d ultrasound based on kernel regression. Biomedical engineering
  online  \textbf{13}(1),  1--15 (2014)

\bibitem{chung2017freehand}
Chung, S.W., Shih, C.C., Huang, C.C.: Freehand three-dimensional ultrasound
  imaging of carotid artery using motion tracking technology. Ultrasonics
  \textbf{74},  11--20 (2017)

\bibitem{daoud2015freehand}
Daoud, M.I., Alshalalfah, A.L., Awwad, F., Al-Najar, M.: Freehand 3d ultrasound
  imaging system using electromagnetic tracking. In: 2015 International
  Conference on Open Source Software Computing (OSSCOM). pp.~1--5. IEEE (2015)

\bibitem{duckelmann2010three}
D{\"u}ckelmann, A.M., Kalache, K.D.: Three-dimensional ultrasound in evaluating
  the fetus. Prenatal diagnosis  \textbf{30}(7),  631--638 (2010)

\bibitem{gonccalves2016three}
Gon{\c{c}}alves, L.F.: Three-dimensional ultrasound of the fetus: how does it
  help? Pediatric radiology  \textbf{46}(2),  177--189 (2016)

\bibitem{kuklisova2012reconstruction}
Kuklisova-Murgasova, M., Quaghebeur, G., Rutherford, M.A., Hajnal, J.V.,
  Schnabel, J.A.: Reconstruction of fetal brain mri with intensity matching and
  complete outlier removal. Medical image analysis  \textbf{16}(8),  1550--1564
  (2012)

\bibitem{mildenhall2020nerf}
Mildenhall, B., Srinivasan, P.P., Tancik, M., Barron, J.T., Ramamoorthi, R.,
  Ng, R.: Nerf: Representing scenes as neural radiance fields for view
  synthesis. In: European Conference on Computer Vision. pp. 405--421. Springer
  (2020)

\bibitem{mohamed2019survey}
Mohamed, F., Siang, C.V.: ‘a survey on 3d ultrasound reconstruction
  techniques. Artificial Intelligence—Applications in Medicine and Biology
  (2019)

\bibitem{moon20163d}
Moon, H., Ju, G., Park, S., Shin, H.: 3d freehand ultrasound reconstruction
  using a piecewise smooth markov random field. Computer Vision and Image
  Understanding  \textbf{151},  101--113 (2016)

\bibitem{moser2019automated}
Moser, F., Huang, R., Papageorghiou, A.T., Papie{\.z}, B.W., Namburete, A.I.:
  Automated fetal brain extraction from clinical ultrasound volumes using 3d
  convolutional neural networks. In: Annual Conference on Medical Image
  Understanding and Analysis. pp. 151--163. Springer (2019)

\bibitem{Papageorghiou2014}
Papageorghiou, A.T., Ohuma, E.O., Altman, D.G., Todros, T., Ismail, L.C.,
  Lambert, A., Jaffer, Y.A., Bertino, E., Gravett, M.G., Purwar, M.:
  International standards for fetal growth based on serial ultrasound
  measurements: the fetal growth longitudinal study of the {INTERGROWTH-21st}
  project. The Lancet  \textbf{384}(9946),  869--879 (2014)

\bibitem{pistorius2010grade}
Pistorius, L., Stoutenbeek, P., Groenendaal, F., De~Vries, L., Manten, G.,
  Mulder, E., Visser, G.: Grade and symmetry of normal fetal cortical
  development: a longitudinal two-and three-dimensional ultrasound study.
  Ultrasound in obstetrics \& gynecology  \textbf{36}(6),  700--708 (2010)

\bibitem{prager2003sensorless}
Prager, R.W., Gee, A.H., Treece, G.M., Cash, C.J., Berman, L.H.: Sensorless
  freehand 3-d ultrasound using regression of the echo intensity. Ultrasound in
  medicine \& biology  \textbf{29}(3),  437--446 (2003)

\bibitem{prevost2017deep}
Prevost, R., Salehi, M., Sprung, J., Ladikos, A., Bauer, R., Wein, W.: Deep
  learning for sensorless 3d freehand ultrasound imaging. In: International
  conference on medical image computing and computer-assisted intervention. pp.
  628--636. Springer (2017)

\bibitem{rahaman2019spectral}
Rahaman, N., Baratin, A., Arpit, D., Draxler, F., Lin, M., Hamprecht, F.,
  Bengio, Y., Courville, A.: On the spectral bias of neural networks. In:
  International Conference on Machine Learning. pp. 5301--5310. PMLR (2019)

\bibitem{sitzmann2020implicit}
Sitzmann, V., Martel, J., Bergman, A., Lindell, D., Wetzstein, G.: Implicit
  neural representations with periodic activation functions. Advances in Neural
  Information Processing Systems  \textbf{33} (2020)

\bibitem{sitzmann2019scene}
Sitzmann, V., Zollh{\"o}fer, M., Wetzstein, G.: Scene representation networks:
  Continuous 3d-structure-aware neural scene representations. arXiv preprint
  arXiv:1906.01618  (2019)

\bibitem{wang2004image}
Wang, Z., Bovik, A.C., Sheikh, H.R., Simoncelli, E.P.: Image quality
  assessment: from error visibility to structural similarity. IEEE transactions
  on image processing  \textbf{13}(4),  600--612 (2004)

\bibitem{wang2021nerfmm}
Wang, Z., Wu, S., Xie, W., Chen, M., Prisacariu, V.A.: Ne{RF}$--$: Neural
  radiance fields without known camera parameters. arXiv preprint
  arXiv:2102.07064  (2021)

\bibitem{yeung2021learning}
Yeung, P.H., Aliasi, M., Papageorghiou, A.T., Haak, M., Xie, W., Namburete,
  A.I.: Learning to map 2d ultrasound images into 3d space with minimal human
  annotation. Medical Image Analysis  \textbf{70},  101998 (2021)

\bibitem{yoo2009fast}
Yoo, J.C., Han, T.H.: Fast normalized cross-correlation. Circuits, systems and
  signal processing  \textbf{28}(6),  819--843 (2009)

\end{thebibliography}


\end{document}